\pdfoutput=1
\documentclass[journal=nalefd,manuscript=letter,layout=twocolumn,]{achemso}
\usepackage[sc]{mathpazo}
\usepackage[scaled=0.9]{helvet}
\usepackage[utf8]{inputenc}
\usepackage{graphicx}
\usepackage{amsmath,amssymb,amsfonts}
\usepackage{upgreek} 
\usepackage{textcomp}
\usepackage[pdftex,dvipsnames]{xcolor}
\usepackage[pdftex]{hyperref}
\hypersetup{
	pdftitle={Near-infrared intersubband photodetection in GaN/AlN nanowires},
	colorlinks,
	citecolor=blue
	}

\title{Near-infrared intersubband photodetection in GaN/AlN nanowires}


\author{Jonas L\"ahnemann}
\affiliation{Universit\'e Grenoble-Alpes, CEA, INAC, PHELIQS, 17 av. des Martyrs, 38000 Grenoble, France}
\alsoaffiliation{Present address: Paul-Drude-Institut für Festkörperelektronik, Leibniz-Institut im Forschungsverbund Berlin e.V., Hausvogteiplatz 5--7, 10117 Berlin, Germany}
\email{jonas.laehnemann@pdi-berlin.de}
\author{Akhil Ajay}
\affiliation{Universit\'e Grenoble-Alpes, CEA, INAC, PHELIQS, 17 av. des Martyrs, 38000 Grenoble, France}
\author{Martien I. Den Hertog}
\affiliation{Universit\'e Grenoble-Alpes, CNRS, Institut Néel, 25 av. des Martyrs, 38000 Grenoble, France}
\author{Eva Monroy}
\affiliation{Universit\'e Grenoble-Alpes, CEA, INAC, PHELIQS, 17 av. des Martyrs, 38000 Grenoble, France}
\email{eva.monroy@cea.fr}

\begin{document}

\begin{abstract}
Intersubband optoelectronic devices rely on transitions between quantum-confined electron levels in semiconductor heterostructures, which enables infrared (IR) photodetection in the 1--30~$\upmu$m wavelength window with picosecond response times. Incorporating nanowires as active media could enable
an independent control over the electrical cross-section of the device and the optical absorption cross-section. Furthermore, the three-dimensional carrier confinement in nanowire heterostructures opens new possibilities to tune the carrier relaxation time. However, the generation of structural defects and the surface sensitivity of GaAs nanowires have so far hindered the fabrication of nanowire intersubband devices. Here, we report the first demonstration of intersubband photodetection in a nanowire, using GaN nanowires containing a GaN/AlN superlattice absorbing at 1.55~$\upmu$m. The combination of spectral photocurrent measurements with 8-band \textbf{k}$\cdot$\textbf{p} calculations of the electronic structure supports the interpretation of the result as intersubband photodetection in these extremely short-period superlattices. We observe a linear dependence of the photocurrent with the incident illumination power, which confirms the insensitivity of the intersubband process to surface states and highlights how architectures featuring large surface-to-volume ratios are suitable as intersubband photodetectors. Our analysis of the photocurrent characteristics points out routes for an improvement of the device performance. This first nanowire based intersubband photodetector represents a technological breakthrough that paves the way to a powerful device platform with potential for ultrafast, ultrasensitive photodetectors and highly-efficient quantum cascade emitters with improved thermal stability.
\end{abstract}

\vspace{10mm}

Using intersubband transitions between quantum-confined electron levels in heterostructures, semiconductors can be exploited for optical devices throughout the infrared (IR) spectral range.\cite{Liu_2000} Well-known examples of commercial intersubband devices are quantum cascade lasers (QCLs)\cite{Faist_1994,Williams_2007,Yao_2012} and quantum well infrared photodetectors (QWIPs).\cite{Levine_1993} In the III-As material system, QWIPs in the 5--30~$\upmu$m wavelength window can achieve picosecond response times, outperforming other semiconductor-based devices in terms of speed.\cite{Levine_1993} III-nitride materials open perspectives for room-temperature operation of intersubband devices both in the near-IR (1--5~$\upmu$m) and in the mid- to far-IR (5--30~$\upmu$m) spectral ranges.\cite{Hofstetter_2010,Beeler_2013,Chen_2015} Even more importantly, the III-nitrides offer the opportunity to integrate nanowires as active media, which represents the ultimate downscaling of devices incorporating quantum well superlattices.\cite{Gudiksen_2002} 

Nanowire geometries are particularly interesting for flexible electronics\cite{Liu_2015} or for the implementation of on-chip optical interconnects.\cite{Brubaker_2013,Tchernycheva_2014a} Photodetectors based on nanowires are characterized by an ultrahigh photocurrent gain ($G\approx 10^5$--$10^8$),\cite{Soci_2007,Cao_2010,Gonzalez-Posada_2012,Lahnemann_2016} though often associated with persistent photoconductivity. Furthermore, nanowires present a large dielectric mismatch with their surroundings, and their diameter is generally smaller than the detected wavelength. These interesting features allow the engineering of the refractive index, or the electrical device cross-section, while maintaining the absorption characteristics of the bulk.\cite{Yan_2009,Krogstrup_2013,Xu_2015,Diedenhofen_2011} The ultrafast relaxation times of intersubband transitions\cite{Tanaka_2008} and their insensitivity to surface phenomena\cite{Beeler_2014} constitute major advantages of nanowire photodetectors. Based on current planar technologies, an obvious material choice for nanowire intersubband devices would be GaAs/AlAs. Unfortunately, the pronounced crystal polytypism in GaAs nanowires obtained through the bottom-up approach\cite{Lehmann_2015} impedes the application as intersubband devices, and the top-down strategy (patterning and etching) appears as the only alternative.\cite{Krall_2015} In contrast, for GaN/AlN, the bottom-up method\cite{Lu_2007} yields the necessary low density of structural defects in spite of the lattice mismatch.\cite{Landre_2010} Thereby, for the III-nitride material system, the nanowire geometry may allow overcoming some of the material quality issues that have hampered the development of III-nitride intersubband devices to date. Planar III-nitride intersubband devices require the use of free-standing GaN substrates to obtain the required low densities of dislocations, and still suffer from crack propagation in structures with high Al content.\cite{Beeler_2013} In contrast, for nanowires, dislocations are restricted to coalescence boundaries,\cite{Consonni_2009} and they can be readily grown on cheap, large-area Si(111) substrates. Reports of intersubband absorption in nanowire ensembles\cite{Tanaka_2008,Beeler_2014,Ajay_2017} and of resonant tunneling transport in single nanowires\cite{Songmuang_2010a,Rigutti_2010} further highlight GaN/AlN heterostructures as promising choice. 

In this study, we demonstrate the feasibility of intersubband photodetection in nanowire-based devices. Using GaN/AlN superlattices embedded in GaN nanowires, intersubband absorption around 1.55~$\upmu$m (telecommunication wavelength) is achieved. An in-depth characterization of the photocurrent characteristics of contacted nanowires under near-IR illumination allows us to unambiguously assign the observed signal to intersubband transitions in the embedded quantum discs. In particular, we observe a linear scaling of the intersubband photocurrent with the incident illumination power, which confirms the insensitivity of the process to surface states.

\begin{figure*}[t]
\includegraphics*[width=16cm]{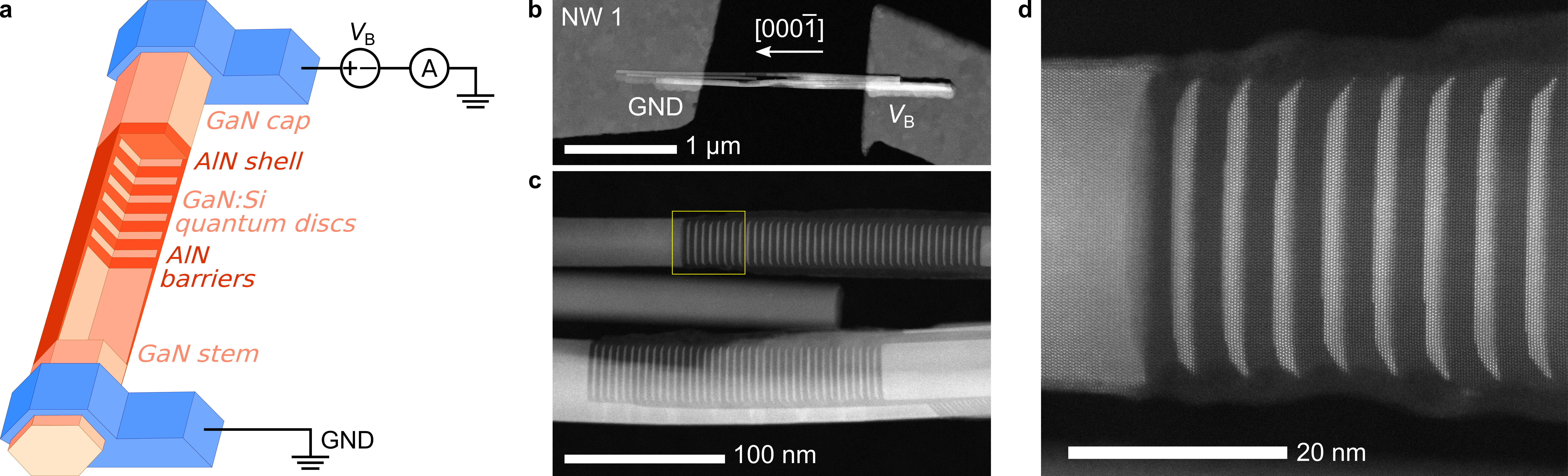}
\caption{\label{fig:tem} (a) Sketch of the investigated nanowire heterostructures, depicting also the contact scheme for photocurrent measurements. (b) Overview HAADF STEM image of NW~1, and (c) detail of the GaN (bright)/AlN (dark) superlattice. (d) High-resolution HAADF STEM image of the region marked in (c) viewed along the  [$2\bar{1}\bar{1}0$] direction. Both the growth direction and the contacting convention are labelled in (b).}
\end{figure*}



To achieve near-IR intersubband photodetection, we investigate GaN nanowires grown by plasma-assisted molecular beam epitaxy (MBE). These nanowires incorporate a superlattice of 39 periods of Si-doped GaN quantum disks separated by AlN barriers and surrounded by an AlN shell, as sketched in Fig.~\ref{fig:tem}a. To populate the ground state in the conduction band, the disks were doped with Si at a doping density of $N_d\approx 3\times10^{19}$~cm$^{-3}$. Nanowires were dispersed on custom-made Si$_3$N$_4$ membranes and contacted by electron beam lithography. Thereby, the very same nanowire studied as a photodetector could be characterized by scanning transmission electron microscopy (STEM). Unless indicated otherwise, the figures in this paper describe the results obtained for one typical specimen (NW 1). However, the study was validated by the observation of similar results in various nanowires with the same structure, as described in the supporting information.

High-angle annular dark field (HAADF) STEM images of NW~1, presented in Figs.~\ref{fig:tem}b--d, demonstrate that the nanowire incorporates regular superlattices with a thickness of $1.6\pm0.3$~nm for the GaN quantum disks and $3.1\pm0.4$~nm for the AlN barriers. The AlN shell, generated by lateral growth when depositing the AlN barriers, exhibits a maximum thickness around 5~nm, which decreases towards the top of the superlattice. Shadow effects during the growth, due to the directionality of MBE, also lead to a reduction of the shell thickness along the nanowire base as we move away from the superlattice. The individual nanowires have a diameter of 30--50~nm. However, in the case of the structure in Fig.~\ref{fig:tem}, due to the coalescence of the nanowire stems, two nanowires with well separated superlattices are contacted in parallel.

\begin{figure*}[t]
\includegraphics*[width=17.5cm]{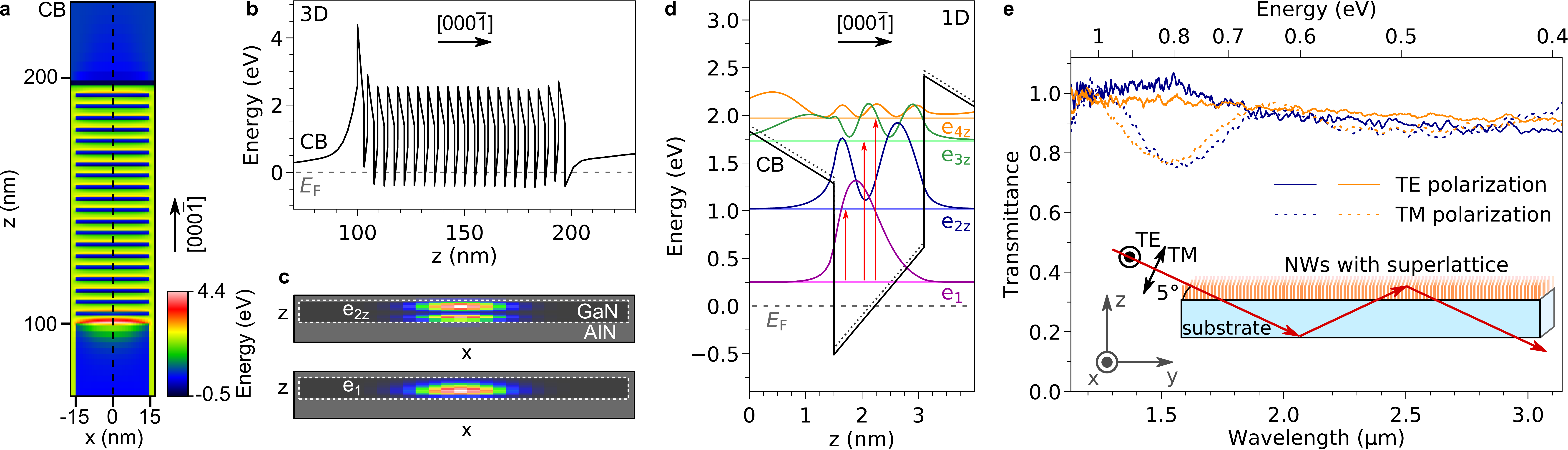}
\caption{\label{fig:sim} (a) Color-coded cross-sectional view of the conduction band (CB) edge as obtained from three-dimensional calculations of the strain and bands in the nanowire heterostructure. (b) Conduction band profile along the [$000\bar{1}$] direction in the center of the nanowire as marked by the dashed line in (a). The grey dashed line indicates the Fermi level ($E_\mathrm{F}$) (c) Cross-sectional view of the spatial extent of the wavefunctions ($|\Psi(x,z)|^2$) for the electron ground state (e$_1$) and the first excited electron state induced by the vertical confinement (e$_{2\mathrm{z}}$) as obtained from \textbf{k}$\cdot$\textbf{p} calculations in the tenth quantum disk. (d) One-dimensional \textbf{k}$\cdot$\textbf{p} calculations of the energy levels and wavefunctions in the quantum disks giving energies of 0.77, 1.4 and 1.71~eV for the e$_1$--e$_{2\mathrm{z}}$, e$_1$--e$_{3\mathrm{z}}$ and e$_1$--e$_{4\mathrm{z}}$ transitions (as marked by the red arrows), respectively. The dotted profile shows the CB edge from the three-dimensional calculation in (b) for comparison. (e) FTIR transmittance of the nanowire ensemble measured for transverse electric (TE, solid) and transverse magnetic (TM, dashed) polarized light at two different positions on the sample. The inset illustrates the measurement geometry with the direction of the electric field indicated both for TM and TE polarization.}
\end{figure*}

Using the dimensions of the quantum disks and barriers extracted from the STEM images, we have performed calculations of the band structure and energy levels in the nanowire superlattice to determine the expected intersubband transition energies. The complex strain distribution in nanowire heterostructures imposes a full three-dimensional analysis. Figure~\ref{fig:sim}a shows a cross-sectional view of the thus calculated energy of the conduction band edge. A profile of the conduction band energy taken along the central axis of the nanowire is depicted in Fig.~\ref{fig:sim}b. It exhibits the polarization-induced sawtooth structure typical for III-nitride quantum wells. Whereas for the lowermost disk, the conduction band lies above the Fermi level, from the 2$^\mathrm{nd}$ disk onward, doping pushes the conduction band below the Fermi level. Note that intersubband absorption is only expected in quantum disks containing electrons in the ground state of the conduction band. We calculated the intersubband transition energy and wavefunctions of the electron ground state (e$_1$) and the first excited electron state associated to the vertical confinement (e$_{2\mathrm{z}}$) in the 10$^\mathrm{th}$ quantum disk. The $(1\bar{1}00)$ cross-sectional views of the squared wavefunctions $|\Psi(x,z)|^2$ (with $x=\langle11\bar{2}0\rangle$ and $z=\langle0001\rangle$) associated to e$_1$ and e$_{2\mathrm{z}}$ are given in Fig.~\ref{fig:sim}c. Both wavefunctions show a maximum probability of finding the electron at the center of the nanowire. Note that this situation, which should facilitate the observation of intersubband absorption, might not be given for other disk dimensions or doping levels.\cite{Beeler_2014} This situation emphasizes the importance of the three-dimensional calculations in combination with a reconstruction of the exact structure by STEM imaging. The e$_1$--e$_{2\mathrm{z}}$ transition energy is 0.76~eV (1.64~$\upmu$m).

Using three-dimensional calculations, the large number of laterally-confined levels prevents us from calculating higher-order levels induced by the vertical confinement. Therefore, we turn to a one-dimensional approximation as depicted in Fig.~\ref{fig:sim}d. This approach is legitimate as the band profile does not deviate significantly from the one extracted from the three-dimensional calculation for the center of the nanowire (dotted line in Fig.~\ref{fig:sim}d). The resulting transition energies for e$_1$--e$_{2\mathrm{z}}$, e$_1$--e$_{3\mathrm{z}}$ and e$_1$--e$_{4\mathrm{z}}$ are 0.77, 1.48 and 1.71~eV (corresponding to wavelengths of 1.61, 0.84 and 0.72~$\upmu$m), respectively. For doped quantum wells, many-body corrections to these transition energies have to be taken into account.\cite{Liu_2000,Beeler_2014} However, for our thin quantum wells and moderate doping densities, the correction is smaller than 10~meV, and will thus be considered as negligible.

To probe the intersubband absorption in the nanowire ensemble, we employed room-temperature Fourier transform infrared (FTIR) spectroscopy at grazing incidence. The selection rules for intersubband transitions require the electric field of the incident light wave to have a component perpendicular to the quantum disk plane, i.e. in the direction where the confined electrons have negligible momentum.\cite{Liu_2000} This condition corresponds to a transverse magnetic (TM) polarization of the light, as indicated in the inset to Fig.~\ref{fig:sim}e. Indeed, the polarization-dependent FTIR transmittance in Fig.~\ref{fig:sim}e exhibits an absorption-related dip between 1.2 and 1.8~$\upmu$m (or 0.7--1~eV) for TM polarized light, while it is constant across the near-IR spectral range for the transverse electric (TE) polarization. The energy range of this absorption feature agrees well with the predicted e$_1$--e$_{2\mathrm{z}}$ transition energy of 0.77~eV, with a relative linewidth $\Delta E/E = 20$\,\%. The broadening and Gaussian shape of the dip, comparable to results obtained in GaN/AlN quantum dots synthesized by the Stranski-Krastanov method,\cite{Guillot_2006} are attributed to the dispersion of the quantum disk size in the nanowire ensemble.

\begin{figure}[t]
\includegraphics*[width=8.5cm]{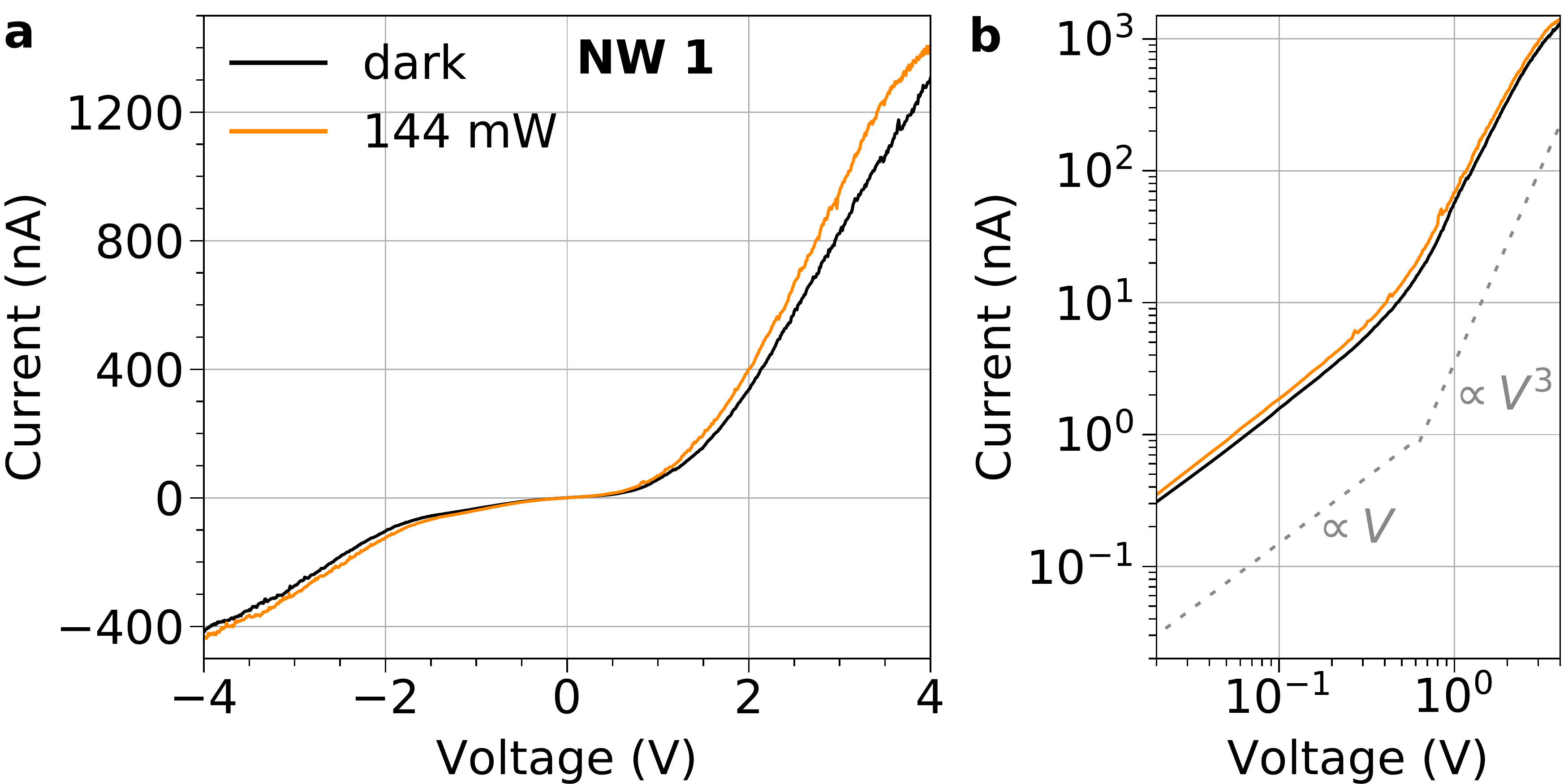}
\caption{\label{fig:iv} Current-voltage characteristics in the dark (black) and under IR laser illumination at 1.55~$\upmu$m (orange) of NW~1 plotted on a linear (left) and logarithmic scale (right). The nanowire was illuminated with a laser power of 144~mW in a spot of 2~mm diameter at the membrane.}
\end{figure}

Having confirmed that the sample under investigation shows intersubband absorption at about 0.8~eV (1.55$~\upmu$m), in agreement with our calculations, we can now turn to the characterization of the IR photocurrent from contacted nanowires. The current-voltage ($I$--$V$) characteristics of NW~1 are presented in Fig.~\ref{fig:iv}. It shows a partially rectifying $I$--$V$ curve with forward (positive) bias being conventionally defined as the direction with higher dark current (indicated in Fig.~\ref{fig:tem}b; see supporting information for a more detailed discussion of the current-voltage characteristics). Plotting the forward current on a log-log scale (Fig.~\ref{fig:iv}b), two major components can be distinguished: a linear ($I\propto V$) regime for low bias (up to 0.1--0.5~V), followed by an increase with the third power of the voltage ($I\propto V^3$), which is an indication of space charge limited transport.\cite{Rose_1955,Gonzalez-Posada_2012}

\begin{figure*}[t]
\includegraphics*[width=13.8cm]{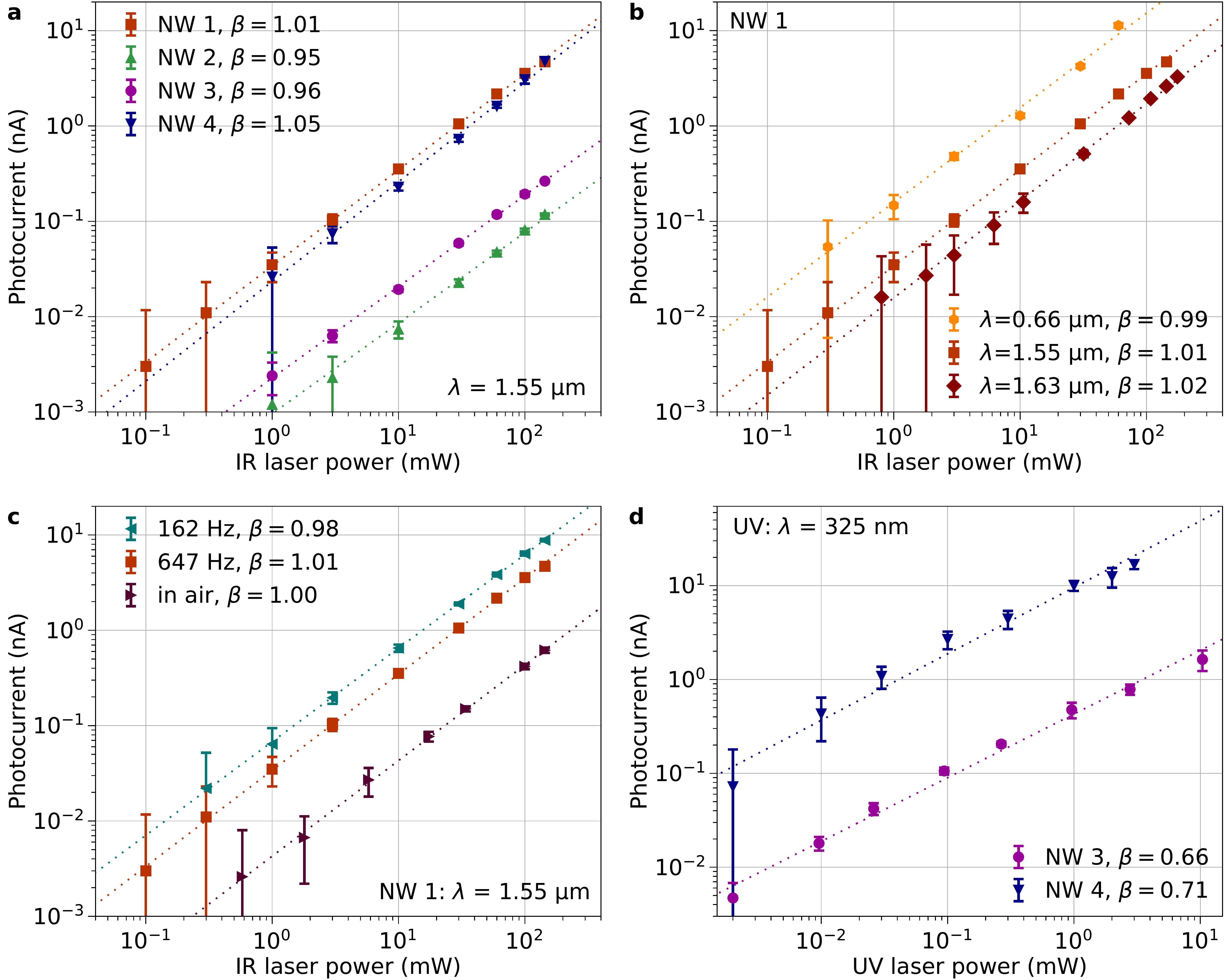}
\caption{\label{fig:lin} Linearity of the intersubband photocurrent for (a) four different nanowires (NWs~1--4) measured in vacuum at an illumination wavelength of 1.55~$\upmu$m (spot diameter at the membrane of 2~mm, laser chopped at 647~Hz), (b) NW~1 measured at different wavelengths (spot diameter at the membrane of 2~mm, lasers chopped at 647~Hz, nanowire in vacuum), as well as (c) NW~1 measured at different chopping frequencies (excitation at 1.55~$\upmu$m, spot diameter at the membrane of 2~mm) in vacuum and in air (laser chopped at 647~Hz). For comparison, (d) gives the photoresponse under UV illumination for NWs~3 and 4 (spot diameter at the membrane of 2~mm, laser chopped at 86~Hz, nanowire in vacuum). The values of $\beta$ given in the legends of the graphs are the power law coefficients determined from fitting the relation $I\propto P^\beta$ to the data. The response to the UV laser is clearly sub-linear, while for the IR illumination, $\beta$ is always close to unity. All measurements are carried out at a bias of 1~V.}
\end{figure*}

Under illumination at 1.55~$\upmu$m, the current increases both in the forward and reverse directions, though more pronounced in the forward direction (the dependence of the photocurrent on the polarization of the incident light is discussed in the supporting information). To remove the contribution from the dark current, we measured the photocurrent under chopped illumination using a lock-in amplifier. The photocurrent as a function of the incident laser power at 1.55~$\upmu$m and under 1~V bias is shown in Fig.~\ref{fig:lin}a for four different nanowires. Strikingly, the photocurrent shows a linear increase over more than three orders of magnitude. The result is confirmed at various IR laser wavelengths, as illustrated in Fig.~\ref{fig:lin}b. Such a linear trend is usually not observed in single nanowire interband photodetectors.\cite{Calarco_2005,Sanford_2010,Gonzalez-Posada_2012,Sanford_2013,Lahnemann_2016} Indeed, the band-to-band photocurrent under ultraviolet (UV) illumination shows a sub-linear dependence on the laser power also for our sample, as demonstrated in Fig.~\ref{fig:lin}d for NWs~3 and 4. This sub-linearity is generally attributed to the importance of surface states, either as non-radiative recombination centers, or as surface charges affecting the nanowire conductivity.\cite{Gonzalez-Posada_2012,Lahnemann_2016} Therefore, the linear power dependence confirms, on the one hand, that IR illumination has no effect on the Fermi level pinning at the nanowire sidewall surfaces, and on the other hand, that the measured photocurrent originates from a mechanism different from the band-to-band photocurrent, with lower sensitivity to surface states. In a previous paper, we have
shown theoretically that the intersubband transition energy is quite robust with respect to surface states and to variations of the strain distribution due to the AlN shell (see Fig.~3b in Ref.~\citenum{Beeler_2014}). 

The linear behavior was reproduced for different chopping frequencies and with the nanowire in air, as shown in Fig.~\ref{fig:lin}c. Note that the photocurrent increases for lower chopping frequencies, which is attributed to the detection electronics (see supporting information). Regarding the reduction of the response when measuring in air, it is explained by the difference in surface state occupation, which modulates the conductivity of the nanowire.

We have estimated the responsivity for NW~1 considering the nanowire surface exposed to the laser as the active photodetector area. Note that the detection cross-section of a nanowire is expected to be larger than this surface.\cite{Xu_2015} We obtain values of $0.6\pm0.1$~A/W and $1.1\pm0.2$~A/W at chopping frequencies of 647 and 162~Hz, respectively (the values are averaged over the different excitation levels in Fig.~\ref{fig:lin}c). To further improve this value, future studies should seek to reduce the nanowire coalescence and the GaN shell growth around the superlattices (see also discussion in the supporting information).

\begin{figure*}[t]
\includegraphics*[width=14cm]{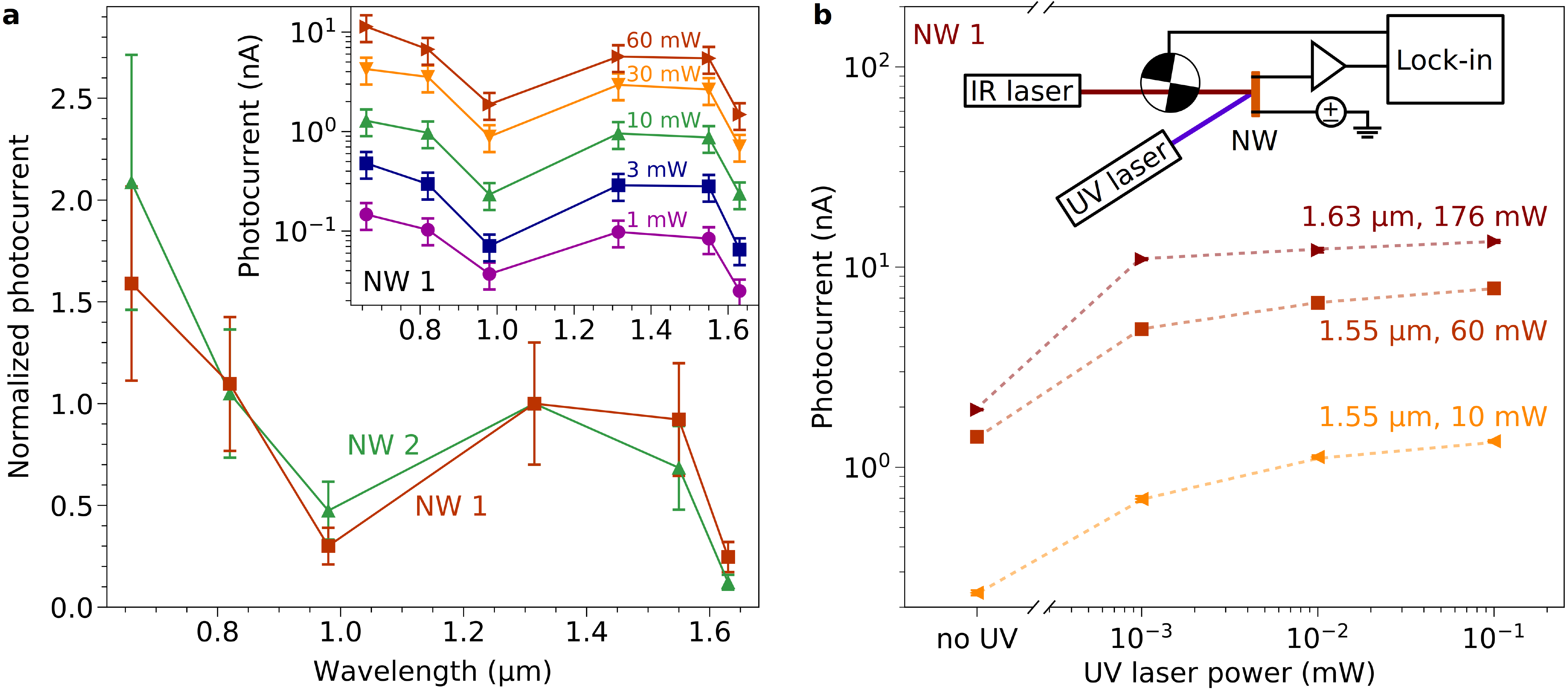}
\caption{\label{fig:resp} (a) Comparison of the normalized near-IR spectral photocurrent response for two nanowires (NW~1 and NW~2) measured at 1~V bias. The response has been averaged over several different illumination powers. The inset shows the spectral response of NW~1 at different illumination levels on a semi-logarithmic scale. The diameter of the laser spot at the membrane was always 2~mm. The error bars account for the uncertainty in the calculation of the impinging irradiance due to the error in estimation of the spot size for the different laser diodes. (b) Evolution of the intersubband photocurrent under additional continuous-wave bias illumination using an UV laser (as sketched in the inset) measured on NW~1. The trend is given for different values of impinging laser power and wavelength. The dashed lines serve as guides to the eye.}
\end{figure*}

The spectral response of the nanowire photodetectors has been obtained by measuring the photocurrent under illumination from laser diodes operating at various wavelengths across the near-IR spectral range, as given in Fig.~\ref{fig:resp}(a) for NW~1 and for another specimen (NW~2). The shape of the spectral response is independent of the incident optical power as shown in the inset to Fig.~\ref{fig:resp}(a). The response reaches a maximum in the 1.3 to 1.55~$\upmu$m range. It goes down around 1.0~$\upmu$m, but then increases again for shorter wavelengths, which can be attributed to transitions from e$_1$ to higher excited states (e$_{3\mathrm{z}}$, e$_{4\mathrm{z}}$), for which also the extraction of the carriers from the quantum disks should be facilitated. The wavelengths corresponding to the e$_1$--e$_{3\mathrm{z}}$ and e$_1$--e$_{4\mathrm{z}}$ transitions were predicted to be 0.84~$\upmu$m and 0.72~$\upmu$m.  Note that transitions between levels with the same parity would be forbidden in square quantum wells.\cite{Liu_2000} However, this restriction is waived in GaN/AlN due to the asymmetry introduced by the internal electric field.

Finally, additional UV illumination should lead to an enhancement of the intersubband photocurrent due to the photogenerated carriers that increase the population of the ground electron level in the conduction band. Indeed, Fig.~\ref{fig:resp}b shows that the intersubband photocurrent from the nanowire increases when adding a continuous-wave UV illumination to the chopped IR illumination, as sketched in the inset. However, a UV laser power of only 1~$\upmu$W improves the measured intersubband photocurrent by a factor 2--3, while further increasing the UV illumination power up to 100~$\upmu$W only leads to a moderate additional increase of the photocurrent. Assuming a carrier lifetime around 10~ns, as measured in nanowire ensembles with similar geometry and doping levels,\cite{Ajay_2017a} and assuming that the impinging laser power is fully absorbed in the quantum disks (unrealistic upper-bound scenario), the number of photogenerated carriers would be $\approx4\times10^{15}$~cm$^{-3}$, which is several orders of magnitude lower than the doping level. Therefore, the increase of the photoresponse in this case is rather due to the effect of the UV illumination on surface states, which shifts the Fermi level higher towards the conduction band, and improves the conductivity of the stem/cap segments,\cite{Gonzalez-Posada_2012,Lahnemann_2016} and in turn enhances the carrier collection. Therefore, while providing further evidence that the IR photocurrent originates from intersubband transitions, this experiment also indicates that a moderate increase of the stem/cap doping, as well as of the quantum disk doping, are pathways for a further improvement of the IR photoresponse.


In conclusion, our work is a proof-of-principle study of intersubband nanowire photodetectors. Near-IR intersubband operation around 1.55~$\upmu$m is achieved using GaN nanowires containing a GaN/AlN superlattice. Unlike the UV band-to-band photocurrent, the IR photocurrent scales linearly with the incident illumination power. This linearity confirms that the UV and IR photocurrents are generated by different mechanisms, the latter being less sensitive to surface-related phenomena, as expected for intersubband transitions in a nanowire heterostructure. As a short-term perspective, the use of ternary (Al,Ga)N barriers would allow to tune the detection wavelength towards the mid-IR spectral region and to enhance extraction of photo-generated carriers. On a broader perspective, our work can be considered a first step towards a nanowire-based intersubband technology, which should lead to ultra-high-speed IR photodetectors with a miniaturization potential down to the scale of tens of nanometers and the possibility to independently tune their electrical (dielectric constant) and optical (absorption) properties. At the same time, the possibility to engineer the electron-phonon coupling in these one-dimensional structures should also pave the way for quantum cascade lasers operating at higher temperatures.

\section*{Methods}

\textbf{Growth} The nanowires investigated in this study were grown on floating-zone Si(111) substrates by plasma-assisted molecular beam epitaxy (MBE) under N-rich conditions\cite{Consonni_2013} at a substrate temperature of $T_S = 810~^\circ$C, and with a growth rate of 330 nm/h. Prior to the growth, the Si(111) substrate was introduced into the MBE chamber and baked at $T_S \approx880~^\circ$C to remove the native oxide. Then, to improve the alignment of the nanowires, an AlN buffer layer was grown prior to the nanowire nucleation, using a two-step procedure, as described in Ref.~\citenum{Ajay_2017a}: The substrate was first cooled down to $T_S = 200~^\circ$C, to deposit 1.2~nm of low-temperature AlN at stoichiometric conditions. Subsequently, the AlN buffer with a total thickness of 8~nm was completed at $T_S = 670~^\circ$C. Following the nanowire nucleation and growth of the 1200~nm long GaN stem at $T_S = 810~^\circ$C, a 39-period GaN/AlN superlattice was formed at the same growth temperature by periodic switching of Ga and Al fluxes. The GaN quantum disks were grown using the same nitrogen-rich conditions that apply to the GaN base, while the AlN sections were grown with an impinging Al flux equivalent to the active nitrogen flux. Finally, the structure was capped with a 1000~nm long GaN segment (see sketch in Fig.~\ref{fig:tem}a). An AlN shell was formed around the superlattice and part of the stem during the growth of the AlN segments, due to the low mobility of the impinging Al atoms.\cite{Furtmayr_2011a} Partially, an additional GaN shell was formed during the cap growth. To facilitate current collection, both the stem and cap, except the regions 25~nm below and 100~nm above the superlattice, were doped n-type with Si to a concentration of $3\times10^{18}$~cm$^{-3}$ (determined from Hall measurements on planar reference samples). To populate the electron ground state in the quantum disks, the GaN disks were Si-doped to a concentration of $3\times10^{19}$~cm$^{-3}$. A scanning electron microscope (SEM) image of the as-grown nanowire ensemble can be found in the supporting information.

\textbf{Processing} To allow photocurrent measurements and scanning transmission electron microscopy (STEM) imaging of the same nanowires, the as-grown nanowires were dispersed on an array of homemade Si$_3$N$_4$ membranes.\cite{denHertog_2012} The nitride membranes have a window size of 200~$\upmu$m and were fabricated from a n$^{++}$ silicon (100) wafer covered on both sides with a SiO$_2$ layer (200~nm) for additional electrical insulation and with a stoichiometric Si$_3$N$_4$ layer (40~nm). Both layers were deposited by low-pressure chemical vapor deposition. Using laser lithography and reactive ion etching, windows and cleavage lines were opened in the Si$_3$N$_4$ and SiO$_2$ layers on one side of the wafer. The etching of the silicon and SiO$_2$ was continued in a KOH bath, leaving only membranes of the top Si$_3$N$_4$ layer. Another optical lithography step, combined with electron beam evaporation of Ti/Au (10~nm / 50~nm) and subsequent lift-off was used to define contact pads and marker structures on the front side of the membrane chips. To disperse nanowires on such membranes, the as-grown sample was sonicated in ethanol, and droplets of the solution were then deposited on the membranes. Contacts to selected nanowires were defined using electron beam lithography, followed by electron beam evaporation of 10~nm of Ti and 120~nm of Al and a lift-off step. See supporting information for SEM images of a Si$_3$N$_4$ membrane and a contacted nanowire.

\textbf{Microscopy} To correlate the photocurrent measurements with the nanowire structure and to determine the superlattice dimensions, the investigated nanowires were imaged 
by high-angle annular dark field (HAADF) STEM in a probe-corrected FEI Titan Themis with a field-emission gun operated at 200 kV. This investigation was carried out only after the full characterization as a photodetector, to avoid any influence of the electron beam exposure on the photocurrent measurements. The thickness of both quantum disks and barriers was determined by averaging over the superlattices visible in Fig.~\ref{fig:tem}c and supporting Fig.~S2 with the error bar corresponding to three times the standard deviation.

\textbf{Simulations} The band structure and transition energies in the nanowire heterostructures were calculated in one and three dimensions with the nextnano3 software\cite{Birner_2007} employing the material parameters for GaN and AlN described in Ref.~\citenum{Kandaswamy_2008}. For three-dimensional calculations, the nanowire was modeled as a hexahedral prism consisting of a 100~nm long GaN section followed by a 20-period AlN/GaN stack and capped with 125~nm of GaN. From the STEM measurements, the geometrical dimensions were defined as follows: radius of the GaN stem of 30~nm, GaN quantum disk thickness of 1.6~nm, AlN barrier thickness of 3.1~nm and AlN shell thickness of 2~nm.  The n-type residual doping was fixed to $3\times10^{17}$~cm$^{-3}$. The stem and cap were n-type doped at a concentration of $3\times10^{18}$~cm$^{-3}$, except a region 25~nm below and 100~nm above the superlattice. The doping of the disks was set to $3\times10^{19}$~cm$^{-3}$. The structure was defined on a GaN substrate to provide a reference in-plane lattice parameter, and was embedded in a rectangular prism of air which allowed elastic strain relaxation. To simulate the effect of surface states, the Fermi level was pinned 0.6 eV below the GaN bandgap at the GaN cap/air interface,\cite{Lymperakis_2013} and 2.1 eV below the AlN band gap at the AlN shell/air interface.\cite{Reddy_2014} The simulation process starts with the calculation of the three-dimensional strain distribution by minimization of the elastic energy assuming zero stress at the nanowire surface. Then, for the calculation of the band profiles, the piezoelectric fields resulting from the strain distribution were taken into account. Wavefunctions and related eigenenergies of the electron and hole states in the quantum disks were obtained by solving the Schr\"odinger-Poisson equations using the 8-band \textbf{k}$\cdot$\textbf{p} model. Additionally, one-dimensional calculations were performed for a segment of the superlattice with periodic boundary conditions. 

\textbf{IR transmittance measurements} The Fourier transform infrared (FTIR) transmittance at room temperature was measured on the nanowire ensemble using a Bruker v70v spectrometer, equipped with a halogen lamp, CaF$_2$ beam splitter, and a nitrogen-cooled mercury-cadmium-telluride detector. The as-grown sample was characterized at a grazing angle of 5$^\circ$ as sketched in the inset to Fig.~\ref{fig:sim}e. The transverse electric (TE) and transverse magnetic (TM) polarizations were distinguished by polarizing the incident light. The floating-zone Si substrate is largely transparent in the investigated spectral range. The transmittance spectra were corrected by a normalization to the transmittance of a nanowire sample without intersubband absorption.

\textbf{Current-voltage characteristics and photocurrent measurements} The current-voltage ($I$--$V$) characteristics were investigated with an Agilent 4155C semiconductor parameter analyzer directly connected to the nanowires. Positive bias is conventionally defined as to have a higher dark current in the forward direction. The end of the nanowire to which the positive bias was applied is indicated in Fig.~\ref{fig:tem}b and supporting Figs.~2a--d. For the measurement of the photocurrent as a function of the optical power, the nanowires were connected to the $10^6$~A/V transimpedance amplifier integrated in a lock-in amplifier (Stanford Research Systems SR830). The laser illumination was chopped at 647~Hz (unless indicated) and the nanowires biased at 1~V. The error bars correspond to three times the standard deviation for measurements averaged over 90~s. To obtain the spectral response, the nanowires were illuminated with different semiconductor lasers operating at wavelengths of 660, 820, 980, 1315, 1550 and 1630~nm. For simultaneous continuous-wave UV illumination above the GaN band gap, a HeCd laser (325~nm) attenuated by optical density filters was focused to the same spot as the IR laser. The same UV laser, chopped at 86~Hz, was used to measure the illumination power-dependent band-to-band photocurrent response. All photocurrent measurements were carried out at room temperature. Unless noted otherwise, the nanowires were held in vacuum during the measurements.

\section{Associated Content}
Supporting information: SEM images of the as-grown nanowire ensemble and of contacted nanowires; additional HAADF STEM characteruzation and correlation to photocurrent measurements for three more nanowires; modulation frequency response of the photocurrent; discussion on the polarization dependence of the intersubband photocurrent.

\begin{acknowledgement}
The authors would like to thank Jean-Michel G\'erard and Oliver Brandt for a critical reading of the manuscript. We benefited from access to the Nanocharacterization platform (PFNC) at CEA Minatec Grenoble. Membrane production and nanowire contacting were carried out at the Nanofab cleanroom of Institut N\'eel, thanks are due to Bruno Fernandez and Jean-Francois Motte for their technical support. This work was financed by the EU ERC Starting Grant TeraGaN (\#278428) and the COSMOS project (ANR-12-JS10-0002) of the French National Research Agency. A.A. acknowledges support from the French National Research Agency via the GaNeX program (ANR-11-LABX-0014).
\end{acknowledgement}

\bibliography{ISB_single_NW_PD_final.bib}

\end{document}